\documentclass[11pt,twoside]{article}

\usepackage{asp2006}
\usepackage{epsf}
\usepackage{psfig}
\usepackage{lscape}

\begin{document}
\title{ ATLAS: DEEP RADIO OBSERVATIONS OF SIX SQUARE DEGREES }   %%% Fill in title
\author{ Ray P. Norris, Enno Middelberg, and Brian J. Boyle }   %%% Fill in author names
\affil{CSIRO Australia Telescope National Facility, PO Box 76, Epping, NSW 1710, Australia}    %%% Fill in author affiliations

\begin{abstract} %%% Abstract to run on from here.

We are using the Australia Telescope Compact Array to image about six square degrees surrounding the Chandra Deep Field South and European Large Area ISO Survey - South 1 regions, with the aim of producing the widest deep radio survey ever attempted, in fields  with deep optical, infrared, and X-ray data. Our goal is to penetrate the heavy dust extinction which is found in active galaxies at all redshifts, and study the star formation activity and active galactic nuclei buried within. Although we are only about half-way through the survey, our data are proving remarkably fruitful. For example, we have discovered a new and unexpected class of object (the Infrared-Faint Radio Sources), we have found that the radio-FIR correlation extends to low flux densities, and we have found powerful AGN-like radio objects in galaxies with a star-forming SED.

\end{abstract}

\section{Introduction}

Deep radio surveys are powerful tools for studying the formation and evolution of galaxies, because they penetrate the dust that can obscure the nucleus, even at infrared wavelengths, and they can unambiguously identify an AGN. Over the last two years, we have conducted the Australia Telescope Large Area Survey (ATLAS) of the Chandra Deep Field South (CDFS) and European Large Area ISO Survey - South 1 (ELAIS-S1) regions, with the aim of producing the widest (6 square degrees) deep (10-15 $\mu$Jy rms) radio survey ever attempted. We have chosen our surveyed areas to coincide with areas imaged by the Spitzer Wide-area Infrared Extragalactic Survey (SWIRE) program (Lonsdale et al. 2003), so that infrared and optical data are available for most of the radio objects. They also encompass the well-studied Great Observatories Origins Deep Survey (GOODS) field in the CDFS (Giavalisco et al., 2004).

There have been a number of very important deep radio surveys (e.g. Condon et al., 2003; Hopkins et al. 2003) which have produced valuable data on radio source statistics, but the potential power of these surveys is often hampered by inadequate data at other wavelengths. Because the ATLAS survey is accompanied by extensive radio, infrared, optical and, in some cases X-ray data, it should make significant advances. 

The key science goals of ATLAS are: 
\begin {itemize}

\item To determine the relative contribution of starbursts and AGN to the overall energy density of the universe, and the relationship between AGN and star-forming activity. Particularly interesting are those cases where the radio AGN lies buried within a host galaxy whose optical/infrared spectrum or SED appears to be that of a star-forming galaxy. It is likely that such sources represent an evolutionary stage in the development of AGNs. 

\item To test whether the radio-far-infrared correlation (RFIC) changes with redshift or with other galaxy properties. Once calibrated, this correlation will be a powerful tool for measuring the star formation history of the Universe. 

\item To search for over-densities of high-z ULIRGs which mark the positions of proto-clusters in the early Universe. With a sampling volume of $ 2 \times  10^{7}$ Mpc$^{3}$deg$^{-2} $ (in the range z = 1-3) this survey will contain at least one proto-cluster with a present-day mass equivalent to Coma, and tens of lower-mass systems. 

\item To trace the radio luminosity function to a high (z $\sim$ 1) redshift for moderate-power sources, and measure for the first time the differential 20 cm source count to a flux density limit of  about 30 $\mu$Jy to a high precision. 

\item To explore a region of parameter space, corresponding to a large area of sky surveyed with high sensitivity at radio, mid-infrared, and far-infrared wavelengths, which would enable us to discover rare but important objects, such as short-lived phases in galaxy evolution.

\end {itemize}

ATLAS focuses on two areas, rather than one, to identify the effect of cosmic variance. Both fields are at least 1.5$\deg$ across, which is sufficient at any redshift to sample structures at least 150 Mpc across at the present epoch. However, Norris et al. (2006) have found that the redshift distribution of radio sources in the CDFS peaks at about z=0.7, presumably because of large-scale structures extending across the field, demonstrating the need to sample widely separate fields.

\section{Observations}

We are currently about half-way through the ATLAS 20 cm survey observations, having covered 6 square degrees of the CDFS and ELAIS-S1 fields to a sensitivity of about 40 $\mu$Jy. When the survey is complete, we hope to reach a final rms of 10-15 $\mu$Jy (depending on time allocation) over this field, and will then release full data products including FITS images and source catalogues. 

In the CDFS field (Norris et al., 2006) we currently reach an rms of 40 $\mu$Jy and identify 784 radio components, which correspond to 726 astrophysically distinct sources, since some sources contain multiple components. In the ELAIS-S1 field (Middelberg et al., 2007) we reach an rms of about 30 $\mu$Jy, and identify 1039 radio components, corresponding to 916 astrophysically distinct sources.

The rms noise level in the CDFS is higher than expected in some parts of the field because of artefacts caused by strong ($\ga$ 1 Jy) sources in the sidelobes of the primary beam. We believe this is caused by the non-circularity of the primary beam pattern, and that it has not been seen before because the ATLAS project is the first to survey such a large area to this depth. We plan to develop software and algorithms to correct for this, because it not only affects ATLAS, but will also be a major challenge for next-generation radio telescopes such as xNTD (CSIRO, 2006) and SKA (Carilli \& Rawlings, 2004). 

In addition to the survey observations themselves, a number of supplementary observational programs are in progress or are being planned, including observations of the entire field at other wavelengths to obtain spectral indices and rotation measures of the polarized objects, Very Long Baseline Interferometry (VLBI) to identify compact AGNs, and spectroscopy.

\begin{figure}[]
\begin{center}
\epsfysize=70mm
\plotone{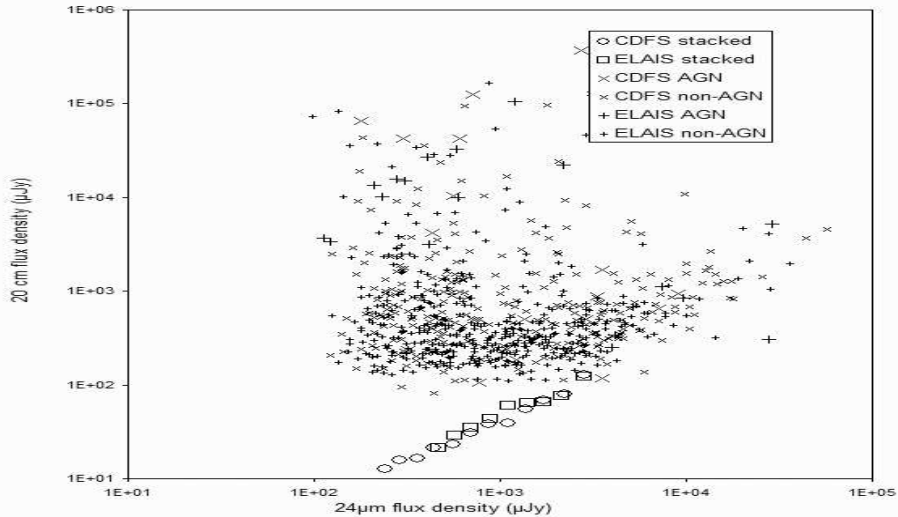 }
%\epsfbox[28 28 567 840]{norris_fig1.eps}
\caption{The observed flux densities of all detected sources in the two ATLAS fields, together with the stacked data from Boyle et al (2007). The \textit{AGN} classification is determined by morphology or spectroscopic classification, while \textit{non-AGN} means that the source has not been classified as an AGN, but may still be either an AGN or star-forming galaxy.}
\end{center}
\end{figure}

\section{The radio-FIR correlation}
It is well-established that a tight correlation exists between the far-infrared and radio emission from low-redshift non-AGN galaxies  (e.g. van der Kruit 1973, Condon et al. 1982, Dickey \& Saltpeter 1984, de Jong et al. 1985). This radio-FIR correlation (RFIC) is believed to be driven by massive stars, which both heat dust, responsible for far-infrared emission, and also, as supernovae, accelerate the cosmic rays that are responsible for the radio synchrotron emission (Harwit \& Pacini, 1975; Condon, 1992; Xu, Lisenfield \& Volk 1994). 

Radio-loud AGNs do not follow the RFIC (e.g. Sopp \& Alexander, 1991), and this fact can be used to separate AGN from SF activity, since a source which departs from the RFIC is likely to be an AGN. However, a source which lies on the RFIC may still be an AGN, because Roy et al. (1998) showed that most Seyfert galaxies also follow it, despite the presence of an AGN. 

By extrapolating observations made with the Infrared Space Observatory (ISO) in the mid-infrared (Cohen et al. 2000, Garrett 2002) and in the sub-millimetre with the SCUBA instrument, (Carilli \& Yun 1999, Ivison et al. 2002), it has been argued that the RFIC extends to z $\ga$ 1. Appleton et al. (2004), have shown a good correlation between the 24 $\mu$m and 20 cm flux for objects with a median z = 0.3 extending out to z $\sim$ 1, and Beswick et al. (2007) show that the luminosities of all identified 24 $\mu$m sources in their extremely deep radio field still follow the correlation for sources with 24 $\mu$m flux densities $\ga$ 80 $\mu$Jy. Exploration of the RFIC at even fainter levels can only be obtained at present by the technique of stacking, and Boyle et al. (2007) have stacked ATLAS radio data to show that the RFIC extends down to microJy levels.

Fig. 1 shows all ATLAS sources, in both the CDFS and ELAIS-S1 fields, which have measured 20 cm and 24 $\mu$m flux densities. These sources include both star-forming galaxies, which presumably follow the RFIC, and AGN, which are expected to lie above it. As a result, a lower bound is clearly visible, presumably corresponding to the RFIC. 

Also shown are the stacked data from Boyle et al. Although the stacked data roughly follow the correlation, it is clear from Fig. 1 that their extrapolation lies below the lower bound to the data. While this might be caused by a systematic error in the Boyle et al stacking procedure, their simulations appear to rule this out. An alternative explanation is that the change in slope is caused by k-correction, although current understanding of the required k-corrections seems to make this unlikely. The remaining explanation is that the weak radio sources represent a different population of objects, or objects with an intrinsically different value of q24, from objects seen in the local Universe. At present, the answer remains unclear, although the higher sensitivity ATLAS data expected in the future will help to resolve the issue.

\section{Infrared-Faint Radio Sources (IFRS)}

Given the relative sensitivities of the ATLAS and SWIRE surveys, we expect that, whether the radio emission is being produced by star formation or by an AGN, all radio sources detected by ATLAS should appear in the SWIRE catalog. Unexpectedly, we find that a small number of radio sources in our sample are not visible at any Spitzer wavelength. We denote this rare class of objects "Infrared-Faint Radio Sources" (IFRS). 

There are 53 such objects in the combined CDFS/ELAIS-S1 sample of 1842 detected radio sources. While the weakest of these may be caused by statistically unusual noise peaks or imaging artifacts, some of them are as strong as 5 mJy, and are clearly not artifacts or noise peaks. Fig. 2 shows two examples, taken from Norris et al. (2006). In both cases, the sources are invisible in all Spitzer infrared wavebands. 

A natural explanation for these objects is that they represent the tail of a distribution of radio/infrared flux densities, and fall just below the Spitzer detection limit. In that case, they should appear in a stacked Spitzer image, even if only a few Spitzer frames are stacked. However, Norris et al (2006) have shown that the stacked Spitzer images do not show a detection, implying that the distribution of infrared fluxes for the radio sources is either bimodal, or else has a very long tail.

We have recently (Norris et al. 2007) detected one of the IFRS's using VLBI, implying that it has an AGN core. It is therefore probably an AGN galaxy so heavily obscured, or at such a high redshift, that all its dust emission is radiated at far-infrared wavelengths beyond 24$\mu$m, and thus undetectable by Spitzer. Alternatively it may be an AGN in a transitory phase in which radio emission is being produced by electrons, but there is insufficient warm dust to produce detectable infrared emission.

\begin{figure}[]
\begin{center}
\plotone{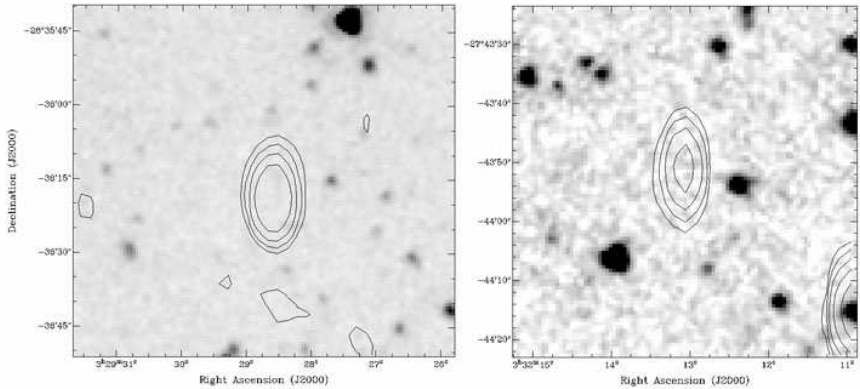}
%\epsfbox[28 28 567 792]{Ha_P.eps}
%\epsfxsize=\hsize\epsfbox{Ha_P.eps}
\caption{Two examples of infrared-faint radio sources, taken from Norris et al. (2006). Contours show 20 cm radio emission, and greyscale shows 3.6 $\mu$m infrared emission, taken from SWIRE data.}
\end{center}
\end{figure}

\section{AGNs and starburst galaxies}

\subsection{Distinguishing AGN from star-formation activity}

There is currently no unambiguous way to decide whether the luminosity of a galaxy is dominated by AGN or star-formation activity. Even Chandra X-ray CDFS observations are too shallow and cover too small an area to be helpful in more than a few cases, while optical and infrared observations do not tell us what is happening in a heavily obscured core

While radio observations cannot give an unambiguous answer, they do supply more clues:
\begin {enumerate}

\item Radio morphology can show unambiguously the familiar double-sided signature of an AGN, but is useful in only a few cases, because of the limited spatial resolution of the ATCA and VLA, and because extended AGN structures are often too weak to be imaged with VLBI. 

\item Spectral indices give us a valuable clue to the type of activity. A spectral index of -0.7 signifies either a star-forming galaxy or a low-power AGN, while high-power AGNs have an integrated spectrum that may either be flat or steep, depending on the level of core domination. For example, of the 13 sources in the HDF-S discussed in detail by Norris et al. (2005), six can be classified as AGN on the basis of the spectral index.

\item The ordered magnetic field of the jets and lobes of an AGN implies that the integrated flux of even a spatially unresolved AGN can contain a significant degree of linear polarization, whilst the tangled magnetic fields of star-forming galaxies tend to produce a negligible integrated polarization. 

\item AGN can depart strongly from the RFIC. However, while a source which departs from the correlation is likely to be an AGN, it cannot be concluded that a source which follows the correlation is not an AGN.

\end{enumerate}

As a result, we expect that this and future deep radio surveys will have an increasing impact on our understanding of the evolution of galaxies in the early Universe.

\begin{figure}[]
\begin{center}
\plotone{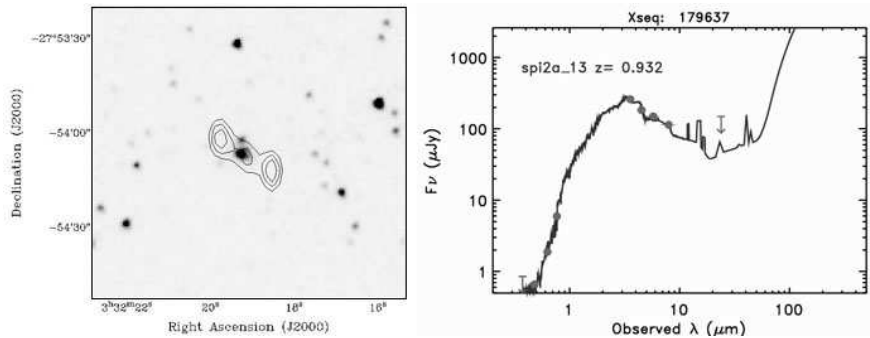}
%\epsfbox[28 28 567 792]{Ha_P.eps}
%\epsfxsize=\hsize\epsfbox{Ha_P.eps}
\caption{The composite source S425. The left image shows radio contours superimposed on a 3.6 $\mu$m SWIRE image, and the right image (Mari Poletta, private communication) shows the SED. This source has the radio morphology of an AGN, but the SED of a star-forming galaxy.}
\end{center}
\end{figure}

\subsection{Obscured AGN}

Because the radio observations penetrate the dust that can obscure the nucleus, they enable us to detect obscured AGN. An example is the source S425, shown in Fig. 3. Its morphology is that of a classic triple radio source, and this is confirmed by its luminosity. With a flux density of 9 mJy at a photometric redshift of 0.932, it has a luminosity of $4 \times 10^{25}$ WHz$^{-1}$, placing it close to the FRI/FRII break, which is consistent with the morphology.

The host galaxy, however, has the SED of a spiral galaxy. Unfortunately it is too faint at optical wavelengths for the Sky Survey images to confirm this classification. The ATLAS catalogue includes several such examples of galaxies whose SED is of a star-forming spiral galaxy, but which have the radio luminosity or morphology of an AGN. We suggest that these represent a class of an AGN buried deeply inside a dusty star-forming galaxy. 

\section{Conclusion and Future Work}

Because of their ability to penetrate dust and reveal AGN characteristics, deep wide radio surveys such as ATLAS and the First Look Survey (Condon et al. 2003) provide information not available at other wavelengths, and are starting to have a profound impact on our understanding of the evolution of high-redshift galaxies. The ATLAS survey is only about half-way through its survey observations, and already significant new results are emerging. So far, however, we have produced more puzzles than answers, including:
\begin {itemize}

\item What are the IFRS?
\item Why does the radio-FIR correlation appear to change slope at faint radio flux densities?
\item What is the significance of the obscured AGN in star-forming galaxies?
\end {itemize}
We hope that, as the ATLAS project continues to develop, it will tackle these and other questions, but we expect that this exploration of a hitherto uncharted region of parameter space will also raise further questions.

\acknowledgements %%% Text of acknowledgements runs on after this command.

The results presented here are the product of a collaboration between the authors and Jose Afonso, Phil Appleton, Stefano Berta,
Paolo Ciliegi, Tim Cornwell,
Scott Croom, Minh  Huynh, Rob Ivison, Carole Jackson, Anton  Koekemoer, Carol  Lonsdale, Bahram Mobasher, Seb Oliver, Mari Polletta, Stefano Rubele,
Brian Siana, Ian Smail, Maxim  Voronkov. We thank them all for their contributions to this work. We particularly thank Mari Poletta for the SED fit shown in Fig. 3.

\end{document}